\documentclass[11pt,a4paper]{article}

\usepackage[cm]{fullpage}
\usepackage{natbib}
\usepackage{apalike}
\usepackage{amssymb}
\usepackage{amsmath}
\usepackage{amsthm}
\usepackage{graphicx}
\usepackage[hidelinks]{hyperref}
\usepackage{txfonts}
\usepackage{authblk}

\newcommand{\aj}{Astron. J.}

\newcommand{\aap}{Astron. Astrophys.}

\newcommand{\icarus}{Icarus}
\newcommand{\mnras}{MNRAS}

\theoremstyle{remark}

\def\be{\begin{equation}}
\def\ee{\end{equation}}
\def\bea{\begin{eqnarray}}
\def\eea{\end{eqnarray}}

\def\d{{\rm d}}

\begin{document}

\title{Basis functions and uniqueness in orbit-averaged population analysis \\ of asteroids}

\author[1]{Mikko Kaasalainen}
\author[2]{Josef \v{D}urech}
\affil[1]{Tampere University, Tampere, Finland}
\affil[2]{Astronomical Institute, Faculty of Mathematics and Physics, Charles University, Prague, Czech Republic\newline
e-mail: durech@sirrah.troja.mff.cuni.cz}
\renewcommand\Affilfont{\itshape\normalsize}
\date{\small\today}

\maketitle

\abstract{
\noindent Time-resolved photometry of asteroids can be used for shape and spin reconstruction. If the number of measurements per asteroid is not sufficient to create a model, the whole data set can be used to reconstruct the distribution of shape elongations and pole latitudes in the population. This is done by reconstructing amplitudes of lightcurves that are estimated from dispersion of points observed at (assumed) constant aspect angle. Here, we formulate orbit-averaged approach where the observable is the orbit-averaged dispersion of brightness.
}

\bigskip

\section{Introduction}
Time-resolved disk-integrated photometry of asteroids is a rich source from which asteroid spin states and shapes can be reconstructed \citep{Kaa.ea:01, Kaa.ea:02c, Dur.ea:15b}. A successful and unique inversion requires enough photometric measurements from different geometries. 

When the amount of data for individual asteroid is not sufficient, an alternative approach can be used -- instead of shape and spin models of individual targets, we can reconstruct directly the distribution of shape elongations and pole latitudes by applying a mathematically simple model of geometrically scattering ellipsoid. This approach was used for the first time by \cite{Sza.Kis:08} followed by \cite{McN.ea:16}. The analytical formulation of the inverse problem was published by \cite{Nor.ea:17} together with a numerical simulation. Then \cite{Cib.ea:18} used this method to Pan-STARRS data.

The method is based on an assumption that a lightcurve amplitude -- that is affected by the aspect angle and shape elongation -- can be estimated from a few brightness data observed at about the same geometry. However, there are data sets that do do have this data cadence. For example, Gaia asteroid photometry in Data Release 2 \citep{Spo.ea:18} can be used for spin/shape reconstruction of individual targets \citep{Dur.Han:18, Cel.ea:19} but the distribution of measurements does not allow to estimate amplitudes from clusters of points.

Instead, \cite{Mom.ea:18} reconstructed the distribution of shapes and pole latitudes by comparing synthetic populations with the observed Gaia data. Here, we formulate a mathematical model to describe orbit-averaged statistics of brightness data and we show that the inverse problem of reconstructing distribution of shape elongations can be solved if we know the distribution of pole latitudes.

\section{Orbit-averaged observables}

The approach here is analogous to that of \cite{Nor.ea:17}. We assume that the disk-integrated brightness of an asteroid can be modelled as that of a geometrically scattering biaxial ellipsoid with semiaxes $a > b = c = 1$. We put $p := b/a = 1/a$. We also assume that the orbit is in the plane of ecliptic. The (unity-scaled) brightness $L$ is given by \citep[eq.~(1) in][]{Nor.ea:17}
\be
L^2=1+(p^2-1)\sin^2\theta\cos^2\phi\,,
\ee
where $\phi$ is the rotation angle and $\theta$ is the aspect angle defined as 
\be
\cos\theta=\sin\beta\cos\Lambda\,,
\ee
where $\beta$ is pole co-latitude in ecliptic coordinates and $\Lambda$ is the relative orbital position longitude, i.e. the difference between the heliocentric ecliptic longitude of the asteroid $\lambda_\mathrm{a}$ and the longitude of the spin axis direction $\lambda$. We assume that observations are always at opposition geometry, so heliocentric longitudes of the asteroid and the observer are the same. 

\begin{figure}[t]
 \centering\includegraphics[width=0.8\textwidth]{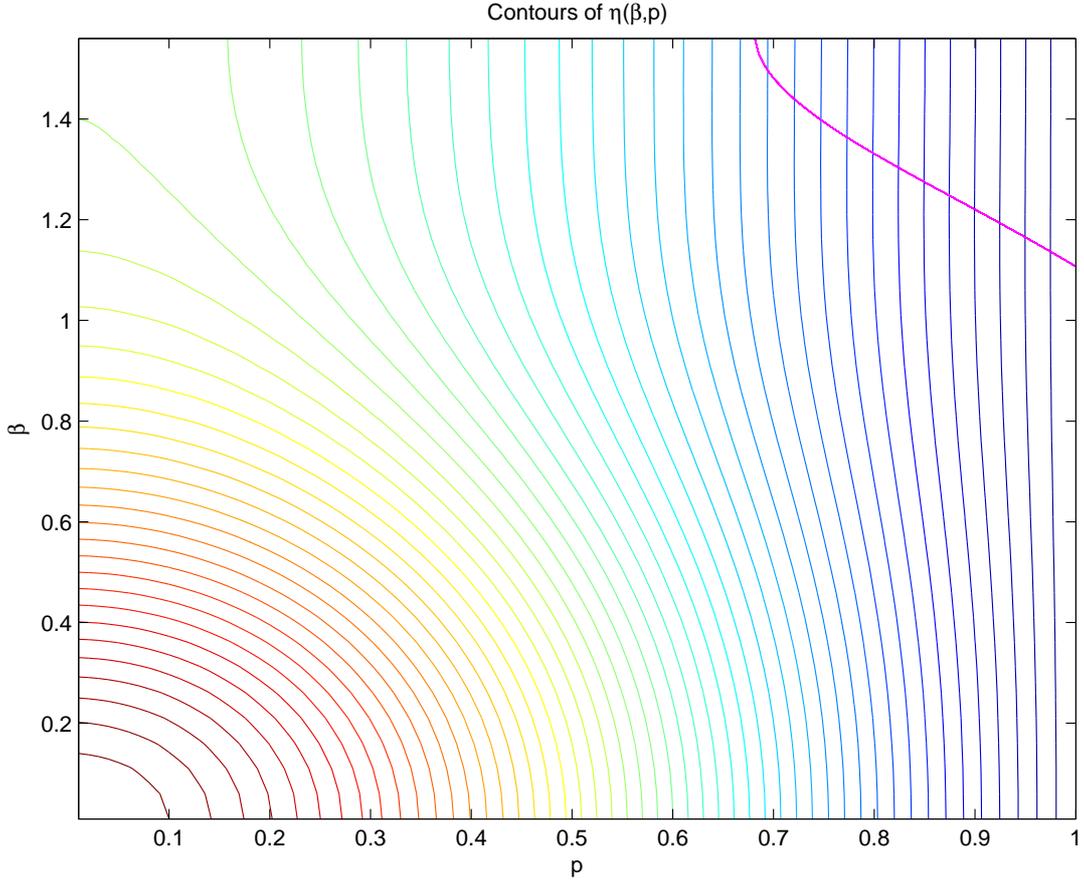}
 \caption{Contours of $\eta$. The magenta curve marks the turning points.}
 \label{fig:eta_contours}
\end{figure}

When averaging over both the rotation $\phi$ and longitude $\Lambda$, we get an orbit-averaged squared brightness
\bea
\left< L^2 \right>_\Lambda & = & \frac{1}{2\pi} \int_0^{2\pi} \frac{1}{2\pi} \int_0^{2\pi}
\left[1 + \left(1 - \sin^2\beta\cos^2\Lambda\right)\left(p^2-1\right)\cos^2\phi\right]\,\d\phi\,\d\Lambda \nonumber \\
&=& 1 + \frac{1}{2}\left(p^2-1\right)\left(1-\frac{\sin^2\beta}{2}\right) \nonumber \\
&=& 1 + \frac{1}{4} \left( p^2 - 1 \right) \left( 1 + \cos^2\beta \right)\,.
\eea
A measure of variation over the average is
\be
\Delta^2 \left(L^2\right)_\Lambda = \left< \left(L^2-\left< L^2\right>_\Lambda\right)^2 \right>_\Lambda
= \left(1-p^2\right)^2 \, \left< \left[\left(1-\sin^2\beta \cos^2\Lambda\right) \, \cos^2\phi - \frac{1}{2}\left(1-\frac{\sin^2\beta}{2}\right)\right]^2\right>_\Lambda\,, 
\ee
where $\left<\cdot\right>_\Lambda$ denotes the averaging double integral over both $\phi$ and $\Lambda$.
This yields
\be
\Delta^2 \left(L^2\right)_\Lambda=\frac{1}{8} \left(1-p^2\right)^2 \, \left(\frac{5}{8}\sin^4\beta-\sin^2\beta+1\right)\,.\label{delta}
\ee
Now we normalize the variation by the mean brightness
\be
\left< L^2 \right>_\Lambda^2 = \frac{1}{4} \left[ \frac{1}{4} \left( p^2-1 \right)^2 \sin^4\beta + \left( 1-p^4 \right) \sin^2\beta + \left( p^2+1 \right)^2 \right]\,.
\ee
Thus, with the dimensionless scale-free normalized dispersion
\be
\label{eq:eta}
\eta^2(\beta,p) = \frac{\Delta^2 \left( L^2 \right)_\Lambda}{\left< L^2 \right>_\Lambda^2}
= \frac{\left( p^2 - 1 \right)^2 \left( \frac{5}{4} \sin^4\beta - 2\sin^2\beta + 2 \right)}{\left( 2p^2 - p^2 \sin^2\beta + \sin^2\beta + 2 \right)^2} = 
\frac{\left( p^2 - 1 \right)^2 \left( \frac{5}{4} \sin^4\beta - 2\sin^2\beta + 2 \right)}{\left[ \left( 1 + \cos^2\beta \right) \left( p^2 + 1 \right) + 2\sin^2\beta \right]^2}\,,
\ee
we have a quadratic equation for $\sin^2\beta$ describing the curves $\beta_\eta(p)$ of constant $\eta$ in the $(p,\beta)$-plane  $p\in[0,1]$, $\beta_\eta(p)\in[0,\pi/2]$:
\be
\sin^4\beta \left[ \frac{\eta^2}{2} \left( p^2-1 \right)^2 - \frac{5}{8} \left( 1-p^2 \right)^2 \right] + \sin^2 \beta \left[ 2\eta^2 \left( 1-p^4 \right) + \left( 1-p^2 \right)^2 \right] + 2\eta^2 \left( p^2+1 \right) - \left( 1-p^2 \right)^2 = 0\,.\label{quadr}
\ee
This has a single feasible root for most of the plane (curves leaning to the left, Fig.~\ref{fig:eta_contours}), but $p$-values on the right half have a turning point at high $\beta$-values; i.e., a feasible double root at which the discriminant $D$ of eq.~(\ref{quadr}) vanishes, after which the curves lean to the right. Technically, $D=0$ occurs because of our choice of observable: the reference average used in eq.~(\ref{delta}) means that $\Delta$ reaches its minimum at $\sin\beta=2/\sqrt 5$ instead of 1, and the normalization of $\eta$ does not alter the non-monotone behaviour for all $p$. This would not be the case if the reference were the local lightcurve average (and the equations would be simpler), but such reference cannot be constructed from sparse data: hence we have to use our $\eta$.

For low values of $\eta$, we have direct information about the shape elongation $p$ because contours of constant $\eta$ are almost vertical lines so the range of possible values of $p$ for a given $\eta$ is narrow even if $\beta$ is not known.

For a given $p$, a possible $\eta$-curve for which the double root lies at the vertical line of $p$ is obtained from the potential feasible root from the condition $D=0$. Hence the value $\beta_\pm$ at which this occurs is obtained by substituting the resulting equation for $\eta^2_\pm(p)$ into the double-root solution
\be
\frac{2\eta_\pm^2 \left( 1-p^4 \right) + \left( 1-p^2 \right)^2}{\frac{1}{2} \eta_\pm^2 \left( p^2-1 \right)^2 - \frac{5}{8} \left( 1-p^2 \right)^2} = -2\sin^2\beta_\pm\,.
\ee
The value $p_0$ after which curves at $\eta<\eta_\pm(p_0)$ have feasible double roots (turning points) is obtained by using $\sin^2\beta_\pm=1$ in the equation above. The curves $\eta>\eta_\pm(p_0)$ have no turning points. The inverse function $p_\pm(\eta)$ for the $p$ at which the curve of given $\eta$ has a double root can also be obtained from the condition equation $D=0$.

The curves of constant $\eta$ start at $\eta=0$ for $p=1$, go through the ``bulge'' emanating from $\sin\beta=2/\sqrt 5$ and moving towards $\beta=\pi/2$ as $p$ decreases until $p_0$ after which the curves lean monotonously to the left. The curves then start to concentrate towards the left corner $(p=0,\ \beta=0)$ where they end at  $\eta=1/\sqrt 2$.

\section{Cumulative distribution functions}
Now we can construct the cumulative distribution function (CDF) of $\eta$ and hence the basis functions for the inverse problem. Obviously the basis functions only pertain to $p$. Information on $\beta$ is lost since a population of a fixed pair of $p,\beta$ gives a single value of $\eta$ that can be caused by infinitely many other pairs, rather than a unique CDF basis function signature as in \cite{Nor.ea:17}. It is thus easiest to fix the $\beta$-distribution and analyse the $p$-distribution.

Let us first construct the CDF $C(\eta)$ for $\eta$ corresponding to curves without turning points, $\eta\ge\eta_\pm(p_0)$. Denoting
\be
1-\sin^2\beta_\eta(p)=\cos^2\beta_\eta(p):=g_\eta(p)\,,
\ee
we have
\bea
C(\eta) & = &\int_{p_{\eta}^{\rm min}}^{p_{\eta}(0)} f(p) \int_{\beta_\eta(p)}^{\pi/2}\sin\beta\, \d\beta\, \d p
+\int_{p_{\eta}(0)}^1 f(p) \int_0^{\pi/2} \sin\beta\, \d\beta \,\d p \nonumber \\
 & = &\int_{p_{\eta}^{\rm min}}^{p_{\eta}(0)} f(p) \sqrt{g_\eta(p)} \,\d p+\int_{p_{\eta}(0)}^1 f(p)\, \d p\,,
\eea
where 
\be
p_{\eta}^{\rm min}=p_{\eta}(\pi/2)\,,\qquad  \sqrt{5}/6\ge\eta\ge\eta_\pm(p_0)
\ee
since the corner $(p=0,\ \beta=\pi/2)$ is met by the curve $\eta=\sqrt{5}/6$, and thus
\be
p_{\eta}^{\rm min}=0\,,\qquad  \eta\ge\sqrt{5}/6\,.
\ee
The basis functions $F_i (\eta)$ for given $p_i\le p_0$ are monotone, 
\be
F_i (\eta)=\sqrt{g_\eta(p_i)}\,,\quad \eta(\pi/2,p_i)\le \eta \le \eta(0,p_i)
\ee
starting at the $\eta$ for the pole in the ecliptic plane:
\be
\eta^2_{\rm min}=\eta^2 \left( \pi/2,p \right) = \frac{5}{16} \frac{ \left( 1-p^2 \right)^2}{ \frac{1}{4}\left( 1-p^2 \right)^2 + 2 \left( p^2+1 \right)}
\ee
before which $F_i=0$, and ending at the $\eta$ for the ecliptically upright pole
\be
\eta^2_{\rm max}=\eta^2(0,p)=\frac{1}{2} \frac{ \left( 1-p^2 \right)^2}{\left( p^2+1 \right)^2}\,,
\ee
after which $F_i=1$ (for the CDF integral, we scan $p$ for a fixed $\eta$, while for $F_i$ we scan $\eta$ w.r.t. a fixed $p$).

The start and end points move monotonically as $p$ grows, so each $p$ has its own unique start and end points,  hence the basis functions are linearly independent. The remaining curves with turning points, $\eta<\eta_\pm(p_0)$, have a slightly more complicated integration geometry, but otherwise the result is the same. Now
\be
C(\eta)=\int_{p_\pm(\eta)}^{p_{\eta}(0)} f(p) \left(\sqrt{g_\eta(p)}-\sqrt{g^+_\eta(p)}\right) \,\d p+\int_{p_{\eta}(0)}^1 f(p)\, \d p\,,
\ee
where $g^+_\eta(p)$ is based on the larger root of the equation (\ref{quadr}) for $\sin^2\beta$. It vanishes at $p_\eta(\pi/2)$, after which we define
\be
g^+_\eta(p)=0\,, \qquad p>p_\eta(\pi/2)\,.
\ee
Thus the basis functions $F_i$ for $p_i>p_0$ are
\be
F_i(\eta)=\sqrt{g_\eta(p_i)}-\sqrt{g^+_\eta(p_i)}\,,
\ee
starting at $\eta_\pm(p_i)$ before which $F_i=0$ and ending at $\eta(0,p_i)$ after which $F_i=1$.

All basis functions have unique start and end points and are linearly independent, so the inverse problem of extracting information on the $p$-distribution from an orbit-averaged observable is uniquely solvable. 

\section{Inversion in practice}
In practice, the question is how well $\Delta(L^2)$ and $\langle L^2\rangle_\Lambda$ are estimated from data not sufficient for sparse photometric analysis. Because there is always noise in data and because the real observing conditions depart from our simple mathematical model, the values of $\eta$ computed from real data will be always overestimated. The main complication is that real observations of main-belt asteroids are not restricted to only exact opposition geometry. They always cover a range of phase angles and the brightness $L$ entering into eq.~(\ref{eq:eta}) has to be corrected for phase angle effects \citep[cf.][]{Mom.ea:18}. The more data we have the better the correction can be but at some point the number of measurements per asteroid will be sufficient for a full inversion \citep{Kaa:04}.

\setlength{\bibhang}{1em}


\newcommand{\SortNoop}[1]{}

\end{document}